\documentstyle[aps,preprint,eqsecnum]{revtex}
\begin{document}
\thispagestyle{empty}
\title{ Gauge Theories: Geometry and cohomological invariants}
\author{M. Kachkachi, A. Lamine and M. Sarih}

\address{
D\'epartement de Math\'ematiques, F.S.T.S.,B.P. 577,\\
Universit\'e Hassan 1$^{\rm er}$, Settat, Morocco}
\maketitle

\begin{abstract}
We develop a geometrical structure of the manifolds $\Gamma$ and $\hat\Gamma$ 
associated respectively to the gauge symmetry and to the BRST symmetry. Then, we
show that ($\hat\Gamma,\hat\zeta,\Gamma$), where $\hat\zeta$ is the group of BRST
transformations, is endowed with the structure of a principle fiber bundle over the
base manifold $\Gamma$. Furthermore, in this geometrical set up due to the
nilpotency of the BRST operator, we prove that the effective action of a gauge theory
is a  BRST-exact term up to the classical action. Then, we conclude that the
effective action where only the gauge symmetry is fixed, is cohomologically
equivalent to the action where the gauge and the BRST symmetries are fixed.
\end{abstract}
\newpage
\section{ Introduction}
The BRST symmetry has been discovered  independently by C. Becchi, A. Rouet and R.
Stora [1] and by I.V. Tyutin [2] as an invariance of the effective Yang-Mills
action. This symmetry is the clue to the proof of renormalizabilty  of a gauge
theory and the starting point for the algebraic determination of chiral anomalies
[3]. Then, it was realized that the BRST invariance in quantum field theories is a
fundamental requirement for a consistent definition of theories with local gauge
invariance[4]. Indeed, in an invariant gauge theory the gauge degrees of freedom are
not physical and must be eliminated. This is done by the usual gauge fixing
procedure in the perturbative Lagrangian appoch. One adds to the gauge invariant
Lagrangian a gauge breaking term rendering the gauge field free propagator well
defined and leading to the decoupling of gauge degrees of freedom from the physical
ones. This decoupling is guaranteed by the requirement of the BRST invariance of the
effective action. Hence, gauge symmetry( and its subsequent BRST gauge fixing) gives
us an unexpected freedom in defining equivalent formulations of the same physical
theory. 

In particular, in their original formulation, topological field theories
were constructed to have the  global symmetry that arises as the BRST symmetry of an
appropriate quantum field theory [5]. This formulation allows for the possibility of
different gauge choices and, these theories are then seen as specific gauge fixings
of a higher theory [6]. So, at the quantum level gauge invariance is
breaking and it is the BRST invariance which takes place. 

By using the nilpotency of the
BRST operator we have shown that the BRST symmetry can not be fixed in any way and
any "BRST fixing condition" is cohomologically equivalent to the gauge fixing one 
[7].

In this paper we develop a geometrical structure of the manifolds
$\Gamma$ and $\hat\Gamma$. Then, we show that the manifold $\Gamma$ is a principal
fiber bundle over the base space-time manifold $M$ whose structural group is the
gauge transformations group. Moreover, we establish that
($\hat\Gamma,\hat\zeta,\Gamma$), where $\hat\zeta$ is the group of the BRST
transformations, is a principal fiber bundle over $\Gamma$. In this
geometrical
formalism we associate gauge fixing conditions $F_{1}(A)=0$ and $F_{2}(A)=0$
respectively to surfaces $\Sigma_{F_{1}}$ and $\Sigma_{F_{2}}$ ( over the manifold
$\Gamma$ ) which are connected by a gauge transformation  and we associate BRST
fixing conditions $G_{1}(A,\Psi)=0$ and $G_{2}(A,\Psi)=0$ respectively to surfaces
$\Sigma_{G_{1}}$ and $\Sigma_{G_{2}}$ (over the manifold  $\hat\Gamma$)
which are connected by a BRST transformation. Furthermore, we establish a
cohomological equivalence of a gauge fixing
condition and a BRST fixing one. Then, we get the cohomological equivalence of the effective action and an "effective action
whose BRST symmetry is fixed". This is interpreted by the fact that the BRST symmetry does not impose any more condition on
the theory and equivalently, it is always possible ( by an appropriate BRST transformation) to go back to the initial gauge
condition. These results allow us to conclude that the effective action is
the sum of the classical action and a BRST-exact term.

The outline of this paper is as follows. In section 2 we give the fiber bundle structures of the manifolds $\Gamma$ and
$\hat\Gamma$ by introducing a principal fiber bundle ($P,G,M$). We show also that a gauge field associated to a connection
over $P$ is a local section of the fiber bundle ($\Gamma,\zeta,M$). By imposing a gauge fixing condition $F(A)=0$, which is
represented by a constant function $F$ on the fibers of the fiber bundle $\Gamma$, the gauge symmetry is fixed.Then, we have
to deal with the effective action that encodes the BRST invariance and to  consider the geometrical structure of  the
manifold $\hat\Gamma$. This enables us to define the principal fiber
bundle ($\hat\Gamma,\hat\zeta,\Gamma$) and to identify
one of its local sections with the BRST operator $Q$ . In section 3, by using this geometrical interpretation of the BRST
operator, we get the equivalence of two BRST fixing conditions defined on the manifold $\hat\Gamma$  and we  prove that a
constraint on $\Gamma$ is cohomologically equivalent to a constraint on $\hat\Gamma$ that is; a $\hat\Gamma$- constraint is
equal to a $\Gamma$-constraint plus a $Q$-commutator term.  This cohomological equivalence enables us to prove that the
BRST fixing action is equivalent to the effective action and hence, the fixation of BRST invariance only adds a
$Q$-commutator term to the effective action. It only changes the gauge fixing term (does not affect the physical degrees
of freedom) and then the BRST symmetry is always present in the effective action. In section 4 we illustrate our results by
the Yang-Mills theories case and we stress that our formalism is general
for any gauge theories. Section 5 is devoted to our
conclusion and open problems.       

\section{ Fibration of the manifolds $\Gamma$ and $\hat\Gamma$}
\subsection{Gauge fields and gauge transformations}
We consider a $G$-principle fiber bundle ($P,G,M$) over the manifold $M$, with $G$ a
compact Lie group, endowed with a local trialization ($U_{i},\varphi_{i}$)
such that, for each $i$, the application $\tau_{i}:M\to P$ given by
$\tau_{i}(p)=\varphi_{i}^{-1}(p,e)$, where $e$ is the unity element of $G$, defines a
local section of the fiber bundle ($P,G,M$). Next, we introduce a
connection $\omega$ on the fiber bundle $P$ as follows:
\begin{equation}
A_{i}=\tau_{i}^{*}\omega,
\end{equation}
where ($A_{i}$) are Lie $G$-valued one-forms over the open sets ($U_i$)
and $\tau_i^*$ is the pullback of $\tau_i$. The one-forms ($A_i$) satisfy
the following relations [6]
\begin{equation}
A_j=\Psi_{ij}^{-1}A_i\Psi_{ij}+\Psi_{ij}^{-1}d\Psi_{ij},
\end{equation}
with $\Psi_{ij}$: $U_i\cap U_j\to G$ are transition functions of the
fiber bundle ($P,G,M$). Reciprocally we have [8]:

{\bf Proposition 1}:

if the one-forms $A_{i}$, which are Lie $G$-valued and defined on
the open set $U_{i}$, verify eq.(2.2) then there exists one and only one connection
$\omega$ satisfying the relation $ A_{i}=\tau_{i}^{*}\omega$ over $U_{i}$.
Consequently we have

{\bf Corollary 1:}

let us consider two local sections $\tau_{1}$ and $\tau_{2}$ such
that 
\begin{equation}
\tau_{2}(p)=\tau_{1}(p)g(p),
\end{equation}
where $g:M\to G$. Then the associated local forms $A_{1}$
and $A_{2}$
verify:
\begin{equation}
A_2 =g^{-1}A_{1}g+g^{-1}dg.
\end{equation}
However, in the geometrical interpretation in terms of the fiber bundle structure,
gauge fields noted $A_{i}$ are viewed as one-forms over the base manifold $M$ and
verify an analogous relationship to eq.(2.4). Moreover, the gauge transformations
group $\zeta$ is the group of vertical automorphisms $\varphi$ over the fiber bundle
($P,G,M$) that satisfy the  base condition
\begin{equation}
\varphi(p)= p 
\end{equation}
for some $p$. In other words, a gauge transformations is an automorphism
which
commutes with the action of $G$.
\begin{equation}
\varphi(gp)= g\varphi(p).
\end{equation}
Locally, gauge transformations are specified by the following proposition [8]:

{\bf Proposition 2:}

A gauge transformation $f$ is completely determined by
a family of applications ($\alpha_{i}$) such that 
\begin{equation}
\alpha_{i}= \varphi_i f \tau_{i},
\end{equation}
where $\varphi_{i} \in C^\infty(U_{i},G)$ and $\alpha_{j}=
ad(\Psi_{ij}^{-1})\alpha_{i}$. $ad$ is the adjoint representation of the
group
$G$ and ($\Psi_{ij}$) are  transition functions of ($P,G,M$). Reciprocally, a family
($\alpha_{i}$) of applications satisfying eq.(2.7) defines a gauge transformation
\begin{equation}
f= (\tau_{i}\pi)[(\alpha_{i}\pi)\varphi_{i}]
\end{equation}
over  $\pi^{-1}(U_{i})$.
Moreover, by expressing explicitly $f^{*}\omega$ we get the gauge
transformation action $f$ on a connection $\omega$  and then, by using the
relation (2.1), we express its action on the gauge field
$A$ as follows [8,9]
\begin{equation}
A'= f^{-1}Af+ f^{-1}df.
\end{equation}
Then, gauge transformations group is realized on the the manifold $M$ as
diffeomorphisms group.

\subsection{The fiber bundle ($\Gamma,\zeta,M$)}
The manifold $\Gamma$ can be equipped with a principal fiber bundle structure over
the manifold $M$ whose orbits are gauge orbits and whose structural group is the
gauge transformations group. Indeed, we define the canonical projection of $\Gamma$
over $M$ by 
\begin{eqnarray}
\tilde\pi &:& \Gamma \to M\nonumber\\
& & A(x) \to \tilde\pi(A(x))= x
\end{eqnarray}
and we consider the gauge field as a map: $M \to \Gamma$ and as a local
section
of the fiber bundle ($\Gamma,\zeta,M$). Equation (2.9) shows also that the action of
$\zeta$ on $\Gamma$ is free ( does not have any fixed point ). Moreover, $\Gamma$ is
locally trivial: let ($U_{i},\varphi_{i}$) be a local trivialization of the fiber
bundle $P$. Then,
the family ($\tilde\pi^{-1}(U_{i}),\tilde\varphi_{i})$ is a local
trivialization of $\Gamma$ such that
\begin{eqnarray}
\tilde\varphi_{i}&:& \tilde\pi^{-1}(U_{i}) \to U_{i}\times\zeta\nonumber\\
                 & & A \to (\tilde\pi(A),\tilde\varphi_{i}(A)),
\end{eqnarray}
where $\tilde\varphi_i (A)$ is the matricial representation of the gauge field in
Lie$G$, i.e.
\begin{equation}
A =A_{\mu}(x)dx^{\mu} =A^{a}_{\mu}(x)T^{a}dx^{\mu}
\end{equation}
and $T^{a}$ is a Lie$G$ basis.
Moreover, we define the transition functions of the fiber bundle
($\Gamma,\zeta,M$) as the
applications
\begin{eqnarray}
\tilde\Psi_{ij}&:& U_{i}\cap U_{j} \to \zeta\nonumber\\
               & &         x \to Ad(\Psi^{-1}_{ij}(x)),
\end{eqnarray}
where $Ad$ is the adjoint representation of Lie$G$. Then, we consider a system
($A_{1},\cdots,A_{n}$) of gauge fields on the manifold $\Gamma$ (which
represents the gauge orbits) such that
\begin{equation}
\forall A\in \Gamma \ \ \exists \ (m_{i}) / A= m_{i}A^{i}
\end{equation}
and hence, ($x_{\mu},m_{i}$) defines a local  coordinates system on
$\Gamma$.

\subsection{ BRST transformations}
The BRST quantization of gauge theories is stated by introducing a gauge fixing term
in the Lagrangian to get the free propagators of gauge fields well defined.This
fixing term implies the presence of non physical degrees of freedom in the theory
but are canceled by ghosts fields. Then the effective action is given by 
\begin{equation}
S_{eff} = S_{0}+ S_{FP}+ S_{GF},
\end{equation} 
where $S_{0}$, $S_{GF}$ and $S_{FP}$ are respectively the classical action, the
gauge fixing action and the ghosts action. However, even the gauge invariance was
fixed at this level, another symmetry of the effective action appears that is; the
BRST symmetry generated by the nilpotent operator $Q$. This operator can  be
decomposed as follows:
\begin{equation}
Q = d+ \delta
\end{equation}
such that $Q^{2} = 0$ and $d^{2} = \delta^{2} = \delta d +d \delta = 0$.
$d$ is the
exterior derivative over the manifold $M$ (or section of $\Gamma$) and $\delta$ is
the restriction of $Q$ to the fibers of $\Gamma$. Furthermore, the
cohomological groups associated to the operators  $d$ and $\delta$ are given
by [7]:
\begin{eqnarray}
H^{0}(d) &=& C^\infty({\Gamma\over P_{0}})\nonumber\\
H_{0}(\delta) &=& ({ker\delta \over Im\delta}) = C^\infty(\Gamma),
\end{eqnarray}
where $P_{0}$ is a gauge orbit. The BRST transformations of collections
($A,\Psi$) are defined as [10]:
\begin{eqnarray}
[Q,A] &=& \Psi\nonumber\\
\left[Q^2,A\right] &=& \{ Q,[Q,A] \} = -D\phi
\end{eqnarray}
with $D\phi$ is the covariant derivative of a gauge parameter $\phi$.
Locally, a BRST transformation is considered as a gauge transformation, where the
gauge parameter is replaced by an anticommuting one. Also, we can see from the
splitting (2.16) of the BRST operator $Q$, that the group of gauge transformations is
a sub-group of the BRST transformations group. We will see in the
paragraph E that
the group of BRST transformations, $\hat\zeta$,is the structural group of the fiber
bundle ($\hat\Gamma,\hat\zeta,\Gamma$).

\subsection{Geometrical interpretation of ghosts fields}
Let us consider a chart $r$ defined by:
\begin{eqnarray}
r&:&U \times V \to P\nonumber\\
& &(x,y) \to r(x,y)= \varphi^{-1}_{i}(a(x), g(y)),
\end{eqnarray},
where $a: M \to M$ and $g: G \to G$ and, let us
consider a connection $\omega$ on the fiber bundle ($P,G,M$) which is expressed on
the chart $r$  as follows:
$$
\omega(x,y) = ad(g(y)^{-1})A_{\mu}(x)dx^{\mu}
+g^{-1}(y)\partial_{\alpha}g(y)dy^{\alpha}.
$$
Its vertical part can be rewritten as:
\begin{eqnarray}
C(y) &=&C_{\alpha}(y)dy^{\alpha}\nonumber\\
C_{\alpha(y)} &=&  g(y)^{-1}\partial_{\alpha}g(y).
\end{eqnarray}
Since $C_{\alpha}(y)\in LieG$, it can be written as $C_{\alpha}(y)
=C^{a}_{\alpha}(y)T^{a}$.
At this level, we identify the usual $F.P.$ fields with the real one-forms $C^{a}(y)
=C^{a}_{\alpha}dy^{\alpha}$ and the ghost field $\Psi$ with its covariant
derivative.
$$
\Psi_{\mu} = - D_{\mu}C.
$$
Hence, due to the fact that the operator $Q$ is the exterior derivative on the
manifold $\Gamma$, the ghost field appears as a $Q$-exact term on the fiber
bundle ($\Gamma,\zeta,M$),i.e. $\Psi = [Q,A]$ which expresses the BRST
transformation of the field $A$.
Here we note that, because
\begin{eqnarray*}
H^{0}(Q) &=& \{ {\rm equivalence \ classes \ of \ physical \
obsevables}\}\\
&=& C^{\infty}\left({\Gamma \over P_{0}}\right) ={ImQ \over kerQ},
\end{eqnarray*}
the introduction of ghosts fields in the theory does not affect the physical
observables.
\subsection{ fiber bundle structure of the manifold $\hat\Gamma$}
As we have noted before,the BRST operator $Q$ is the exterior derivative on the
manifold $\Gamma$ and can be written locally in the form
$$
Q = dm_{i} {\partial \over \partial m_{i}} 
$$
\begin{equation}
[Q,m_{i}] = dm_{i}.
\end{equation}
The manifold $\hat\Gamma$ is defined as the set of all collections
($A,\Psi$) (over which the effective action is BRST-invariant) and then can
be endowed with a fiber bundle structure.

{\bf Proposition 3:}

The manifold $\hat\Gamma$ is a principal fiber bundle over the manifold
$\Gamma$ whose structural group is  the BRST transformations group. Then, the
canonical projection of $\hat\Gamma$ over $\Gamma$ is defined by
\begin{eqnarray}
\hat\Pi &:& \hat\Gamma \to \Gamma\nonumber\\
          & & (A,\Psi) \to \hat\Pi(A,\Psi) = A.
\end{eqnarray}
Also, $Q$ and $\Psi$ are local sections of the fiber bundle
($\hat\Gamma,\hat\zeta,\Gamma$). Indeed, the local trivialization of $\hat\Gamma$ is
defined by the family
($\hat\Pi^{-1}(\tilde\Pi^{-1}(U_{i})),\hat\varphi_i$) such that
\begin{eqnarray}
\hat\varphi_{i} &:& \hat\Pi^{-1}(\tilde\Pi^{-1}(U_{i})) \to
\tilde\Pi^{-1}(U_{i})\times\hat\zeta\nonumber\\
                & &(A,\Psi)  \to (A,\hat\varphi_{i}(\Psi)),
\end{eqnarray}
where $\hat\varphi(\Psi)$ is the matricial representation of the ghost field in the
Lie algebra of the group  $\zeta$. Then, the transition functions of the fiber bundle
$\hat\Gamma$,denoted by $\hat\Psi_{ij}$, are defined as follows:
\begin{eqnarray}
\hat\Psi_{ij} &:& \tilde\Pi^{-1}(U_{i}) \cap \tilde\Pi^{-1}(U_{j}) \to
\zeta\nonumber\\
               & & A \to ad(\hat\Psi_{ij}(A)),
\end{eqnarray}
where $ad$ is the adjoint representation of the group $\zeta$          
\section{ Cohomological equivalence between $\Gamma$ and $\hat\Gamma$}
In the geometrical language the quantization of gauge theories, which is done at the
first step  by breaking the gauge symmetry, is equivalent to choosing only one
representative from each gauge equivalent classes, i.e. picking out a point from
each gauge orbit and to  consider a constraint-surface $\Sigma_{F}$  on the
manifold
$\Gamma$ defined by the equation
\begin{equation}
F(A) = 0,
\end{equation}
with $F$ a map :$\Gamma \to LieG$. This is equivalent to choose a local section of
the fiber bundle($\Gamma,\zeta,M$) which is in our setting a gauge fixing
condition. This implies that the surface $\Sigma_{F}$ meets every orbit in one and only
one point. This assertion induces the following lemma

{\bf Lemma 1}:

All gauge fixing conditions are equivalent and then two constraint-surfaces are
related by a gauge transformation. Indeed, let us consider $\Sigma_{F_{1}}$ and
$\Sigma_{F_{2}}$. Since $F_{1}$ and $F_{2}$ are local sections of $\Gamma$ then there
exists an element of the structural group $\zeta$ of the fiber bundle ($\Gamma,\zeta,M$)
connecting the two surfaces,i.e. $\exists f \in \zeta / \Sigma_{F_{2}} = f^{*}
\Sigma_{F_{1}}$.

In the same way, we define a BRST fixing condition on $ \hat\Gamma$ by the equation
\begin{equation}
G(A,\Psi) = 0,
\end{equation}
Where $G: \hat\Gamma \to Lie\hat\zeta$. This is associated to a local
section on $\hat\Gamma$. Furthermore, we associate to the eq.(3.2) a $G$-constraint
surface $\hat\Sigma_{G}$ defined on the manifold $\hat\Gamma$.Then, using the fibration
($\hat\Gamma,\hat\zeta,\Gamma$) we have the following lemma:

{\bf Lemma 2}:

Let us consider a map $G(m_{i},\hat m_{i})$ over $\hat\Gamma$. It is constant on the
BRST orbits. Then $G(m_{i}, \hat m_{i})$ takes the form
\begin{equation}
G(m_{i},\hat m_{i}) = G_{0}(m_{i}) + G_{1}(m_{i})\hat m_{i}  +...,
\end{equation}
Where $G_{0}$ and $G_{1}$ are maps on $\Gamma$. Indeed, since $G(m_{i},\hat m_{i})$
is constant on the BRST fibers then it projects into a map on $\Gamma$, i.e.
\begin{equation}
\hat\Pi(G(m_{i},\hat m_{i})) = G_{0}(m_{i})
\end{equation}
 and it decomposes into $G_{0}$ and a term depending only on orbits parameters, say
$\hat m_{i}$. So, we have $G(m_{i},\hat m_{i}) = G_{0}(m_{i}) +
G_{1}(m_{i})\hat m_{i}$.
Finally eq.(3.3) and the nilpotency property of the operator $Q$ enable us to get
the following proposition.

{\bf Proposition 4}:

A BRST fixing condition on the manifold $\hat\Gamma$ is equivalent to a gauge fixing
condition on the manifold $\Gamma$ up to a $Q$-exact term. Furthermore, if
$G(A,\Psi)$ is a BRST fixing condition then it takes the form
\begin{equation}
G(A,\Psi) = G_{0}(A) + \{Q,\Lambda\},
\end{equation}
where $\Lambda$ is an arbitrary function of the field $\Psi$.
Indeed,the action of the operator $Q$ on the eq.(3.5) expresses as
\begin{equation}
\{Q,G(A,\Psi)\} = \{Q,G_{0}(A)\} + \{Q,G_{1}(A)\Psi\}
\end{equation}
Whereas,
\begin{eqnarray}
\{Q,G_{1}(A)\Psi\} &=& G_{1}(A)\{Q,\Psi\} +\{Q,G_{1}(A)\}\Psi\nonumber\\
                 &=& G_{1}(A)\{Q,[Q,A]\} +{\partial G_{1}(A)\over\partial A}\Psi^{2}
\end{eqnarray}
The coefficient of $G_{1}$ of the first term (in the last equation ) is the Jacobi
identity which is equal to zero. Furthermore, $\Psi^{2} =0$ (because $\Psi$ is an
anticommuting parameter) and then we have
\begin{equation}
\{Q,G(A,\Psi)\} = \{Q,G_{0}(A)\}.
\end{equation}
So, we conclude that $G(A,\Psi)$ and $G_{0}(A)$ are equivalent up to a $Q$-exact
term that is; $G(A,\Psi) = G_{0}(A) + \{Q,\Lambda\}$. This means  that the $F$-constraint and the $G$-constraint defined respectively on $\Gamma$ and on $\hat\Gamma$ are
cohomologically equivalent and that the projection of the
$G$-constraint surface on the manifold $\Gamma$ gives a $F$-constraint surface :
\begin{equation}
\hat\Pi(\hat\Sigma_{G}) = \Sigma_{F}.
\end{equation}
Otherwise, the splitting of the BRST operator: $Q = d + \delta$ implies
the following proposition

{\bf Proposition 5}:

The effective action is cohomologically equivalent to the classical
action:
\begin{equation}
S_{eff} = S_{0} + \{Q,\Lambda\}.
\end{equation}
Indeed the decomposed expression of $Q$ gives
\begin{equation}
\{Q,S_{0}\} = dS_{0}.
\end{equation}
Furthermore, the gauge invariance of the classical action implies that
\begin{equation}
\{ Q,S_{0} \} = 0
\end{equation}
However, the BRST invariance is expressed as
\begin{equation}
\{ Q, S_{eff} \} = 0.
\end{equation}
 Consequently, we have the following corollary

 {\bf Corollary 2}:

All  the effective actions are equivalent.Indeed, if $S'_{eff}$ is an
other effective action then, eq.(3.10) implies that
\begin{equation}
S'_{eff} = S_{0} + \{ Q,\Lambda'\}.
\end{equation}
and we have
\begin{eqnarray}
S'_{eff} &=& S_{eff} + \{ Q, \Lambda' - \Lambda \}\nonumber\\
         &=& S_{eff} +\{ Q, \Lambda''\}.
\end{eqnarray}
Finally, this algebraic treatment effectively shows that the BRST symmetry can
not be fixed at the quantum level and only the gauge symmetry that can be
fixed.
\section{ Yang-Mills theory}
The effective action associated to the classical action of
a free Yang-Mills theory, when considering a gauge fixing condition
$F(A) = 0$, is given by [11]:
\begin{equation}
S_{eff} = \int d^4x (L_{class} - {1\over2\lambda}F^{a}F_a +
\varpi^a(QF^a)).
\end{equation}

Let $G(A,\Psi)=0$ be a BRST fixing condition, then eq.(3.5) gives
\begin{equation}
G(A,\Psi)=F(A)+\{Q,\Lambda\}
\end{equation}
or,
\begin{equation}
G(A,\Psi)=G_0(A)+G_1(A)\Psi.
\end{equation}
Then replacing the expression of the $F$-constraint given by eqs.(4.2,3)
in eq.(4.1) we get:
\begin{eqnarray}
S_{eff} &=& \int d^4x \left[L_{class} -
{1\over2\lambda}(G^a(A,\Psi)-\{Q,\Lambda^a\})(G_a(A,\Psi)-
\{Q,\Lambda_a\})\right.\nonumber\\
& &\left.+\varpi^a(QG_a(A,\Psi)-\{Q,\Lambda_a\})\right]\nonumber\\
&=&\int d^4x \left[L_{class} -
{1\over2\lambda}G^aG_a+\varpi^a\{Q,G_a\}+{1\over2\lambda}(G^a\{Q,\Lambda_\}
+\{Q,\Lambda^a\}G_a)\right.\nonumber\\
& &\left.-\varpi^a\{Q,\{Q,\Lambda_a\}\}+
{1\over2\lambda}\{Q,\Lambda^a\}\{Q,\Lambda_a\}\right].
\end{eqnarray}
Furthermore, the nilpotency of the operator $Q$ implies that
$\{Q,\{Q,\Lambda\}\}=0$ and $\{Q,\Lambda\}\{Q,\Lambda\}=
\Psi{\partial\Lambda\over\partial A}\Psi{\partial\Lambda\over\partial A}
=\left({\partial\Lambda\over\partial A}\right)^2\Psi^2=0$ since $Q=\Psi
{\partial\over\partial A}$. Otherwise,
\begin{eqnarray}
G(A,\Psi)\{Q,\Lambda\}+\{Q,\Lambda\}G(A,\Psi)&=&(G_0(A)+G_1(A))
\Psi{\partial\Lambda\over\partial A}+\Psi{\partial\Lambda\over\partial A}
(G_0(A)+G_1(A)\Psi)\nonumber\\
&=&G_0\Psi{\partial\Lambda\over\partial A}+\Psi{\partial\Lambda\over\partial A}
G_0\nonumber\\
&=&G_0\{Q,\Lambda\}\nonumber\\
&=&\{Q,\Lambda'\}.
\end{eqnarray}
Then, $S'_{eff}$ is expressed in terms of $S_{eff}$ as follows:
\begin{equation}
S_{eff}=S'_{eff}+\{Q,\Lambda'\},
\end{equation}
where
\begin{equation}
S'_{eff}
=\int d^4x \left[L_{class} -
{1\over2\lambda}G^aG_a+\omega^a\{Q,G_a\}\right]
\end{equation}

\section{Conclusion and perspectives}
In this paper we have established explicitly the fiber bundle structure
of the manifolds $\Gamma$ and $\hat\Gamma$ associated respectively to the
gauge symmetry and to the BRST symmetry. Furthermore, we have shown that
the gauge fixing term (on $\Gamma$) and the BRST fixing term (on
$\hat\Gamma)$ are cohomologically equivalent. Hence the projection of the
second condition on $\Gamma$ gives the first one. This enables us to get
the effective action as the sum of the classical action and of a $Q$-
exact term.

One can try to extend this formalism to topological field theories and
then consider the metric tensor on the two manifolds.

\newpage
\section*{References}
\noindent \ [1] C. Becchi, A. Rouet and R. Stora, Phys. Lett. {\bf 52B} (1974) 344
and Ann. Phys. {\bf 98}

(1978) 287.

\noindent \ [2] I.V. Tyutin, Report Fian. {\bf 39} (1975) unpublished.

\noindent \ [3] C. Les Houches, 1983, B. de Witt and R. Stora eds (North Holland,
1984).

\noindent \ [4] L. Baulieu, Phys. Rep. {\bf 129C} (1985) 1.

\noindent \ [5] E. Witten, J. Diff. Geom. {\bf 17} (1982) 661.

\noindent \ [6] A.S. Schwarz, Lett. Math. Phys. {\bf 2} (1978) 247.

\noindent \ [7] H. Kachkachi and M. Kachkachi, J. Math. Phys. {\bf 35} (1994) 4467.

\noindent \ [8] M. Nakahara, {\it Geometry, Topology and Physics}, IOP Publishing.

\noindent \ [9] D. Birmingham, M. Blau, M. Rakowski and G. Thompson, Phys. Rep.
{\bf 209} (1991) 129.

\noindent [10] E. Witten, Trieste Conference on topological methods in quantum
field theories, Trieste

Italy, 1990 edited by W. Nahm, S. Randjbar -
Daemi, E. Sezgin and E. Witten

(World Scientific, Singapore, 1991).

\noindent [11] F. Gieres, {\it Geometry of Supersymmetric Gauge Theories},
Lectures Notes in Physics 302

Springer - Verlag.

\end{document}